\documentclass[12pt]{iopart}
\usepackage{iopams}
\usepackage{graphics}
\usepackage[next]{inputenc}
\usepackage[dvips]{epsfig}
\usepackage{epsfig,color}
\def\be{\begin{equation}}
\def\ee{\end{equation}}

\def\bi{\begin{itemize}}
\def\ei{\end{itemize}}
\def\bn{\begin{enumerate}}
\def\en{\end{enumerate}}
\def\bea{\begin{eqnarray}}
\def\eea{\end{eqnarray}}

\def\ba{\begin{array}}
\def\ea{\end{array}}
\def\bd{\begin{displaymath}}
\def\ed{\end{displaymath}}
\def\la{\langle}
\def\ra{\rangle}

\begin{document}

\title{Magnetic properties of the spin $S=1/2$ Heisenberg chain with hexamer
modulation of exchange}

\author{M. Shahri Naseri$^1$, G. I. Japaridze$^{2,3}$, S.  Mahdavifar$^1$ and \\S. Farjami Shayesteh$^1$}
\address{$^{1}$ Department of Physics, University of Guilan, 41335-1914, Rasht, Iran}
\address{$^{2}$College of Engineering, Ilia State University, Cholokashvili Ave. 3-5, 0162
Tbilisi, Georgia}
\address{$^{3}$Andronikashvili Institute of Physics, Tamarashvili str. 6, 0177 Tbilisi, Georgia}

\begin{abstract}

\leftskip 1cm \rightskip 1cm

 We consider the spin-1/2 Heisenberg chain with
alternating spin exchange 
in the presence of additional modulation of exchange on odd bonds
with period three. We study the ground state magnetic phase
diagram of this hexamer spin chain in the limit of very strong
antiferromagnetic (AF) exchange on odd bonds using the numerical
Lanczos method and bosonization approach. In the limit of strong
magnetic field commensurate with the dominating AF exchange, the
model is mapped onto an effective $XXZ$ Heisenberg chain in the
presence of uniform and spatially modulated fields, which is
studied using the standard continuum-limit bosonization approach.
In absence of additional hexamer modulation, the model undergoes a
quantum phase transition from a gapped phase into the only
one gapless L\"{u}ttinger liquid (LL) phase by increasing the
magnetic field. In the presence of hexamer modulation,  two new
gapped phases are identified in the ground state  at magnetization
equal to $\frac{1}{3}$ and $\frac{2}{3}$ of the saturation value.
These phases reveal themselves also in magnetization curve as
plateaus at corresponding values of magnetization. As the result,
the magnetic phase diagram of the hexamer chain shows seven
different quantum phases, four gapped and three gapless and the
system is characterized by six critical fields which mark quantum
phase transitions between the ordered gapped and the LL gapless
phases.

\end{abstract}


\pacs{75.10.Jm, 75.10.Pq}

\maketitle
\section{Introduction} \label{sec1}

Plateaus observed in the zero-temperature magnetization curve of
spin systems, usually display the quantum nature of this
phenomenon. Formation of a plateau is connected with generation of
a gap in the excitation spectrum, which can be in some senses
regarded as a  realization of generation the Haldane
conjecture~\cite{haldane83}. The quantum nature of a plateau
formation mechanism was clearly shown in the seminal paper by
Oshikawa, Yamanaka and Affleck \cite{oshikawa97}, who proposed the
condition for realization of a plateau at magnetization $m$ as
$n(S-m)=integer$, where $S$ is the magnitude of the local spin,
$n$ is the number of spins in a translational unit cell of the
ground state and $m$ is normalized to 1 at saturation.

A particular realization of such scenario appears in the
one-dimensional (1D) space-modulated (alternating) quantum spin
systems. The bond alternating Heisenberg spin-1/2 chains which are
obtained by a space modulation in the exchange couplings represent
one particular subclass of low-dimensional quantum magnets which
pose interesting theoretical [3-17]~
and experimental [18-26] problems.
The bond alternating spin-1/2 chains have a gap in the spin
excitation spectrum and reveal extremely rich quantum behaviors in
the presence of an external magnetic field. By turning the
magnetic field, the excitation gap reduces and reaches to zero at
the first critical field. Simultaneously, the magnetization
remains zero up to the first critical field. By more increasing of
the magnetic field, system remains in gapless phase and at the
second critical field, the gap re-opens and the saturation plateau
is appeared in the magnetization curve. These models have only two
plateaus at zero and saturation values of the magnetization.

The subject of the space-modulated spin chains has grown in recent
years and has been found that a space modulation in the exchange
couplings can affect on the behaviors of the field induced
magnetization. First, the modulation was suggested as the
Ferromagnetic-Ferromagnetic-Antiferromagnetic (F-F-AF) trimerized
Heisenberg spin-1/2 chains and found that the magnetization curve
has a plateau at $1/3$ of the saturation magnetization
value~\cite{hida-0-94,okamoto96}. The F-F-AF trimerized chain has
been observed experimentally in the compound
$3CuCl_2.2dx$~\cite{ajiro94}. During the recent years, the
trimerized Heisenberg chains have been studied in
details~\cite{gu05,gu06,gong08}. Using the density matrix
renormalization group (DMRG) method, the observation of the
plateaus for chains with different spin ($S=1/2, 1, 3/2, 2$) has
been reported~\cite{gu05}. A magnetization plateau at $m=1/3$ of
the saturation value exists at low temperature for both, F-F-AF
and AF-AF-F trimerized spin-1/2 chains~\cite{gu06}. The spin
structure factors are also calculated for the trimerized cases and
found that the static structure factor dose not vary with the
external magnetic field at the plateau state~\cite{gong08}.

The other example of the spin chain with mixed (F-AF) and
spatially modulated exchange is the compound $Cu(3-Clpy)_2(N_3)_2$
where $(3-Clpy)=3-Chloroyridine$, which is known as a typical
example of a spin-$\frac{1}{2}$ F-F-AF-AF tetrameric spin
chain~\cite{escuer98}. In this  system, there is a gap from the
singlet ground state to the triplet excited states in the absence
of a magnetic field. The thermodynamic properties of the
tetrameric spin chains with alternating F-F-AF-AF exchange
interactions have been investigated
numerically~\cite{lu05,gong09}. The existence of a plateau at
$m=1/2$ of the saturation value has been observed. The temperature
dependence of the magnetization, susceptibility and specific heat
are studied to characterize the corresponding phases. In recent
paper, one of the authors considered a different tetrameric spin
chain as AF$_{1}$-F-AF$_{2}$-F~\cite{Mahdavifar11}. By means of
numerical Lanczos method, nonlinear sigma model and bosonization
approach, it has been found that a magnetization plateau appears
at $\frac{1}{2}$ of the saturation value. The effects of the space
modulation are reflected in the emergence of a plateau in other
physical functions such as the F-dimer and the bond dimer order
parameters, and the pair-wise entanglement.

In this paper,  we consider the spin-$\frac12$ Heisenberg chains
with alternating exchange supplemented by the additional modulation
of the one subset of bonds with period three, what gives the hexamer
modulation of the spin exchange in the chain.  Beside
the very rich quantum magnetic phase diagram,  this
model has the ability to exhibit two new plateaus theoretically, at
values $1/3$ and $2/3$ of the saturation. We restrict our
consideration by the case where the exchange on the modulated
subset of bonds is antiferromagnetic and substantially larger then
the exchange on undistorted bonds. In the presence of strong
magnetic field commensurate with dominant AF exchange, the model is
mapped onto an effective spin-1/2 XXZ Heisenberg chain in the
presence of both longitudinal uniform and spatially-modulated
fields. This mapping allows us to use the continuum-limit
bosonization analysis and study the ground state phase diagram of
the effective spin-chain model.  We show that the very presence of
additional modulation leads to the dynamical generation of two new
energy scale in the system and to
 the appearance of four additional quantum phase
transitions in the magnetic ground state phase diagram. These
transitions manifest themselves most clearly in the presence of
two new magnetization plateaus at magnetization equal to
$\frac{1}{3}$ and $\frac{2}{3}$ of the saturation value. Also, the
width of the new plateaus are estimated using the scaling
properties of the effective theory.~

We confirm the results obtained within the continuum-limit studies
of the effective model by using the exact-diagonalization of
finite chains and perform  an accurate simulation
 at zero temperature
using the numerical Lanczos method. There are two additional
gapped phases and corresponding two magnetization plateaus in the
magnetization curve of our model at values of magnetization equal
to $\frac{1}{3}$ and $\frac{2}{3}$ of the saturation value. Using
the numerical technique, we also study the magnetic field effects
on the bond-dimer order parameter and string correlation function.
Finally, treating the non-modulated weak bonds as perturbation, we
obtain perfect analytical expressions for parameters
characterizing width of the two additional plateaus.

The paper is organized as follows. In the forthcoming section, we
introduce our alternating space-modulated  spin-1/2 model and in
the strong modulated coupling limit derive an effective spin chain
Hamiltonian in section III. In section IV, we present our
analytical bosonization results. In section V,  the results of a
numerical simulation are presented. Finally, we discuss and
summarize our results in section VI.  In the Appendix, the details
of calculations considering the used perturbation approach are
presented.

\begin{figure}
\centerline{\psfig{file=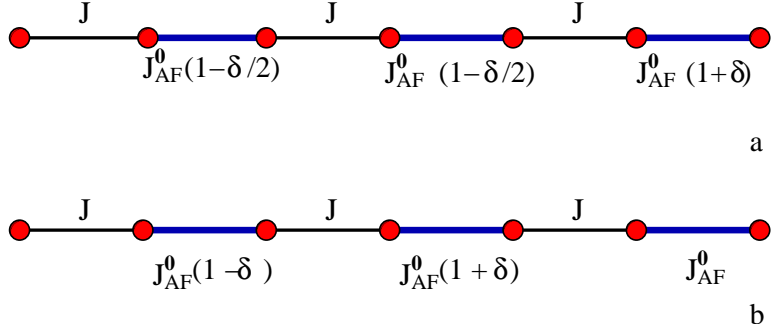,width=4.00in}}
 \caption{(Color online.) Schematic representation of
 spin chains with the hexameric modulation of spin exchange which is considered in the paper.
(a) corresponds to the hexamer distortion of the exchange pattern
"A" type and (b) shows the "B" type.}\label{Schematic}
\end{figure}


\section{The model}     \label{sec2}

The Hamiltonian of the model under consideration on a periodic
chain of N site is defined as
\begin{eqnarray}
{\cal H}&=&J\sum_{n=1}^{N/2}\textbf{S}_{2n}.\textbf{S}_{2n+1}
+\sum_{n=1}^{N/2}J_{AF}(n)\textbf{S}_{2n-1}.\textbf{S}_{2n} -H
\sum_{n=1}^{N/2} \left(S_{2n-1}^{z}+S_{2n}^{z}\right),
\label{Hamiltonian}
\end{eqnarray}
where $S_n$ is the spin-1/2 operator on the $n$-th site, $J>0$ and $J<0$
denote the antiferromagnetic and ferromagnetic couplings respectively. $H$ is the
uniform magnetic field and $J_{AF}(n)$ is
spatially modulated antiferromagnetic exchange. We restrict our
consideration by following two types of antiferromagnetic exchange
modulation corresponding to the hexameric distortion of the exchange
pattern: the $"A"$ type
\begin{eqnarray}
J_{AF}(n) &=& J_{AF}^{0}(1+\delta\cos(\frac{2\pi}{3}n)),
\end{eqnarray}
which corresponds to the hexamer modulation of the spin exchanges
given in Fig.~\ref{Schematic}(a) and the $"B"$ type
\begin{eqnarray}
J_{AF}(n) &=&
J_{AF}^{0}(1-\frac{2\delta}{\sqrt{3}}\sin(\frac{2\pi}{3}n)),
\end{eqnarray}
which corresponds to the following pattern hexamer modulation of
the spin exchanges given in Fig.~\ref{Schematic}(b). The uniform
spin-exchange coupling between spins separated by odd bonds is
considered as $|J| \ll J_{AF}^{0}$.  For $\delta=0$ and $J<0$ the
Hamiltonian (\ref{Hamiltonian}) reduces to the Hamiltonian of
alternating F-AF spin-1/2 chains  and for $J>0$ to the Hamiltonian
of alternating AF-AF spin chains in a longitudinal uniform
magnetic field.


\section{The effective Hamiltonian}

In the considered limiting case of strong antiferromagnetic
exchange on odd bonds $J^{0}_{AF}\gg |J|$ and for magnetic field
$H \simeq J^{0}_{AF}$, one can use the standard
procedure~\cite{mila98, Totsuka98} to map the model
(\ref{Hamiltonian}) onto the effective $XXZ$ model. The easiest
way to obtain the effective model, is to start from the limit
$J=0$, where at $H=0$ the system reduces to the set of
disconnected pairs of spins in singlet state, located on odd bonds
and coupled with strong AF exchange $J_{\small AF}(n)$. At $H \neq
0$ spins on each bond either remain in a singlet state, $|s\ra$,
with energy $E_{s}(n) = -0.75~J_{\small AF}(n)$ or in one of the
triplet states, $|t^{+}\ra$, $|t^{0}\ra$ and $|t^{-}\ra$ with
energies $E_{t^{0}}(n) = 0.25~J_{\small AF}(n)$ and
$E_{t^{\pm}}(n) = 0.25~J_{\small AF}(n) \mp H$, respectively.
As the field $H$ increases, the energy of the triplet
state $|t^{+}\rangle$, decreases and at $H \simeq J_{AF}(n)$
forms, together with the singlet state, a doublet of almost
degenerate low energy state, split from the remaining high energy
two triplet states. This allows to introduce the effective spin
operator $\tau$ which act on these states as
\cite{mila98, mahdavifar08}
\begin{eqnarray}
&&\tau_{n}^{z}|\,s>_{n}~ = -\frac{1}{2}|\,s>_{n}\, , ~~~~~
\tau_{n}^{z}|\,t^{+}>_{n} ~ = \frac{1}{2}|t^{+}>_{n}\, ,\nonumber \\
&&\tau_{n}^{+}|\,s>_{n} ~ = ~~~|\,t^{+}>_{n}\, , ~~~~~
\tau_{n}^{+}|t^{+}>_{n}~ = ~~~0 \, , \\
&&\tau_{n}^{-}|\,s>_{n} ~ = ~~~~ 0 \, ,~~~~~~~~~~~~
\tau_{n}^{-}|\,t^{+}>_{n}~ = ~|\,s>_{n}\,  . \nonumber
\end{eqnarray}
The relation between the real spin operator $\textbf{S}_{n}$ and the
pseudo-spin operator ${\mbox{\boldmath $\tau$}}_{n}$ in this
restricted subspace can be easily derived by inspection,
\begin{eqnarray}
S^{\pm}_{2n-1} &=& - S^{\pm}_{2n} =
\frac{1}{\sqrt{2}}\tau^{\pm}_{n},\,\nonumber \\
S^{z}_{2n-1} & = & S^{z}_{2n}=
\frac{1}{2}\left(\frac{1}{2}+\tau^{z}_{n}\right) \, .
\label{S-Tau-relations} \end{eqnarray}
Using (\ref{S-Tau-relations}), to the first order and up to a
constant, we easily obtain the effective Hamiltonian
\begin{eqnarray}
\hspace{-10mm}H_{eff}&=&
\frac{J}{2}\sum_{n=1}^{N/2}\left[\left(\tau_{n}^{x}\tau_{n+1}^{x}+\tau_{n}^{y}\tau_{n+1}^{y}\right)
+\frac{1}{2}\tau_{n}^{z}\tau_{n+1}^{z}\right] - \sum_{n=1}^{N/2}\,
\left[h_{eff}^{0}+h_{eff}^{1}(n)\right]\tau_{n}^{z}\, ,
\label{effevtive-hamiltoni}
\end{eqnarray}
where
\begin{eqnarray}
h_{eff}^{0} & = & H - J^{0}_{\small AF} - J/4\, , \label{eff}
\end{eqnarray}
and
\begin{eqnarray}
h_{eff}^{1}(n) &=& -\delta J^{0}_{\small AF}\cos(2\pi n/3) \equiv
- h_{1}^{A}\cos(2\pi n/3), \label{h_1_A}
\end{eqnarray}
in the case of "A" type of modulation and
\begin{eqnarray}
h_{eff}^{1}(n) &=& \frac{\textcolor[rgb]{0.00,0.00,0.00}{2}\delta
J^{0}_{\small AF}}{\sqrt{3}}\sin(\frac{2\pi}{3}n)\equiv
h_{1}^{B}\sin(2\pi n/3), \label{h_1_B}
\end{eqnarray}
in the case of "B" type of modulation.

Thus, the effective Hamiltonian is nothing but the anisotropic XXZ
Heisenberg chain in an uniform and spatially trimer modulated
magnetic fields. The anisotropy is $\Delta=1/2$ ($\Delta=-1/2$) in
the case of chain with antiferromagnetic (ferromagnetic) weak
bonds $J>0$  ($J<0$). It is worth to notice that a similar problem
has been studied intensively in past years~\cite{Alcaraz95,
Okamoto96, Essler09}. At $\delta=0$, the effective Hamiltonian
reduces to the XXZ Heisenberg chain in an uniform longitudinal
magnetic field. The magnetization curve of this model has only
saturation plateau corresponding to the fully polarized chain. At
$H=0$ and $J^{0}_{AF} \gg |J|$ spins coupled by strong bonds form
singlet pairs and the singlet ground state of the initial
spin-chain system is well described by superposition of singlets
located on strong bonds with magnetization per bond $M=0$. In
terms of effective $\tau$-spin model, this state corresponds to
the ferromagnetic order with magnetization per-site equal to its
negative saturation value $m=-1/2$. In the opposite limit of very
strong magnetic field $H \gg J^{0}_{\small AF}$, fully polarized
state of the initial chain with magnetization per strong bond,
$M=1$, is represented in terms of effective $\tau$-spin chain as
fully polarized  state with magnetization per site $m=1/2$. This
gives following relation between the magnetization per strong bond
of the initial spin $S=1/2$ chain, $(M)$, and the magnetization
per site of the effective $\tau$-chain (m): $M=m+1/2$.

At $\delta \neq 0$, the effective model corresponds to the
Heisenberg chain in spatially modulated fields with period three.
In this case, in agreement with standard theoretical predictions,
additional magnetization plateaus would appear at values of
magnetization $\frac{1}{3}$ and $\frac{2}{3}$ of the saturation
value. Below, in this paper we use the analytical and numerical
tools to analyze the magnetic phase diagram of the model under
consideration and determine critical values of magnetic fields
corresponding to transitions between sectors of the ground state
phase diagram characterized by different magnetic behavior.

\section{Analytical results} \label{sec3}

\subsection{The magnetization onset critical field $H_{c}^{-}$ and the saturation field $H_{c}^{+}$}

The performed mapping allows to determine critical fields
$H_{c}^{-}$ corresponding to the onset of magnetization in the
system and the saturation field $H_{c}^{+}$ \cite{mila98}. The
easiest way to get $H_{c}^{-}$ and $H_{c}^{+}$ is to perform the
Jordan-Wigner transformation~\cite{Peschel75} which maps the problem
onto a system of interacting spinless fermions with trimerized
modulated on-site potential:
\begin{eqnarray}
\hspace{-10mm}H_{sf} & = &\pm\frac{|J|}{2}
\sum_{n}^{N/2}\left(a_{n}^{\dagger}a^{\phantom{\dagger}}_{n+1} +
h.c.\right)\textcolor[rgb]{0.00,0.00,0.00}{
+\frac{|J|}{4}}\sum_{n}^{N/2}\rho_{n}\rho_{n+1} - \left(\mu_{0} +
\mu_{1}(n)\right)\rho_{n}\, , \label{Hamiltonian_SpFrm}
\end{eqnarray}
where $\rho_{n}=a_{n}^{+}a^{\phantom{+}}_{n}$, 
$\mu_{0} = \textcolor[rgb]{0.00,0.00,0.00}{h_{eff}^{0}} - J/4 = H
- J^{0}_{\small AF}- J/2$, $\mu_{1}(n)=
\textcolor[rgb]{0.00,0.00,0.00}{h_{eff}^{1}}(n)$ and the sign $+$
($-$) corresponds to the $J>0$ ($J<0$).

The lowest critical field $H^{-}_{c}$ corresponds to that value of
the chemical potential $\mu_{0}^{c}$ for which the band of
spinless fermions starts to fill up. In this limit, we can neglect
the interaction term in Eq. (\ref{Hamiltonian_SpFrm}) and obtain
the model of free particles with three band spectrum. Below, in
this subsection we consider only the case of "A" type of exchange
modulation with $\mu_{1}(n)= h_{eff}^{1}(n)$ given by Eq.
(\ref{h_1_A}). Generalization of this results for case "B" is
straightforward.

In this case three bands of the single particle spectrum are given
by
\begin{eqnarray}
E_{1}(k) & = &-H + J^{0}_{\small AF} +J\sqrt{1
+\textcolor[rgb]{0.00,0.00,0.00}{\Delta}^{2} }\cos\varphi(k)\, ,\label{Eigenstates_1}\\
E_{2}(k) & = &-H + J^{0}_{\small AF} + J\sqrt{1
+\textcolor[rgb]{0.00,0.00,0.00}{\Delta}^{2}}\cos\left(\varphi(k)+2\pi/3\right)\, ,\label{Eigenstates_2}\\
E_{3}(k) & = & - H + J^{0}_{\small AF} + J\sqrt{1 +
\textcolor[rgb]{0.00,0.00,0.00}{\Delta}^{2}}\cos\left(\varphi(k)+4\pi/3\right)\,
, \label{Eigenstates_3}
\end{eqnarray}
where  $\textcolor[rgb]{0.00,0.00,0.00}{\Delta}  = \delta
J_{AF}^{0}/J$,
\begin{eqnarray}
\varphi(k)& = &\frac{1}{3}
\arccos\left(\frac{\cos(3k)+\textcolor[rgb]{0.00,0.00,0.00}{\Delta}^{3}}{\sqrt{\left(1+
\textcolor[rgb]{0.00,0.00,0.00}{\Delta}^{2}\right)^{3}}}\right)\,
,
\end{eqnarray}
and $-\pi/3 < k \leq \pi/3$. This gives
\begin{eqnarray}
H_{c}^{-} & = & J^{0}_{\small AF} + J\sqrt{1 +
\textcolor[rgb]{0.00,0.00,0.00}{\Delta}^{2}}\cos\left(\varphi(\pi/3)+4\pi/3\right)\,
\qquad {\mbox at} \qquad J>0\, ,\label{H-c-Minus-AF} \\ H_{c}^{-} &
= & J^{0}_{\small AF} - J\sqrt{1 +
\textcolor[rgb]{0.00,0.00,0.00}{\textcolor[rgb]{0.00,0.00,0.00}{\Delta}}^{2}}\cos\varphi(0)\,
\hspace{ 2.2 cm}\qquad {\mbox at} \qquad J<0\, .\label{H-c-Minus-F}
\end{eqnarray}

To estimate the critical magnetic field $H_{c}^{+}$, which marks the
transition into the phase with saturated magnetization, it is useful
to make a site-dependent particle-hole transformation on the
Hamiltonian of Eq.(\ref{Hamiltonian_SpFrm}): $a_{n}^{\dagger}
\rightarrow d^{\phantom{\dagger}}_{n}$. Up to constant the new
Hamiltonian reads
\begin{eqnarray}
\hspace{-10mm}H_{hole} & = &
\mp\frac{|J|}{2}\sum_{n}^{N/2}\left(d_{n}^{\dagger}d^{\phantom{\dagger}}_{n+1}
+ h.c.\right) +
\frac{J}{4}\sum_{n}^{N/2}\rho^{d}_{n}\rho^{d}_{n+1} -
\left(\mu^{h}_{0} + \mu_{1}(n)\right)\rho^{d}_{n}\, ,
\label{Hamiltonian_SpFrm-NEW}
\end{eqnarray}
where the hole chemical potential $\mu^{h}_{0}=-\mu_{0}+J/2$. In
terms of holes, $H_{c}^{+}$ corresponds to the chemical potential
where band starts to fill up, and one can neglect again the
interaction term. However, the effect of interaction is now
included in the shifted value of the chemical potential for holes.
After simple transformations, we obtain
\begin{eqnarray}
H_{c}^{+} & = & J^{0}_{\small AF} + \frac{J}{2} + J\sqrt{1 +
\textcolor[rgb]{0.00,0.00,0.00}{\gamma}^{2}}\cos\varphi(0)
\hspace{ 1.8 cm}\qquad {\mbox at} \quad J>0\,
,\label{H-c-Minus-AF}\\
H_{c}^{+} & = & J^{0}_{\small AF} + \frac{J}{2} - J\sqrt{1 +
\textcolor[rgb]{0.00,0.00,0.00}{\Delta}^{2}}\cos\left(\varphi(\pi/3)+4\pi/3\right)\,
\quad {\mbox at} \quad J<0\, .\label{H-c-Minus-AF}
\end{eqnarray}

 The spectrum of the system in the free fermion limit
(\ref{Eigenstates_1})-(\ref{Eigenstates_3}) allows to determine two
other important parameters which characterize the values of magnetization in the magnetization curve of the system which the
additional plateaus appear and the values of magnetic field which
correspond to the center of each plateau. Below we consider only the
case $J>0$, however extension to the case $J<0$ is straightforward.

At 1/3-rd band-filling, all states in the lower band $E_{3}(k)$ are
filled and separated from the empty at $E_{2}(k)$ band by the energy
 $2m_{0}=E_{2}(k=0)-E_{3}(0)$. Therefore, the first magnetization
plateau will appear at magnetization equal to $1/3$ of the
saturation value. The magnetic field at the center of plateau
is given by
\begin{eqnarray}
H_{c1}^{0} & = &J^{0}_{\small AF}+ E_{3}(0)+ m_{0}.
 \label{H-c1-0}
\end{eqnarray}
Analogically at 2/3-rd band-filling, all states in the lower bands
$E_{3}(k)$ and $E_{2}(k)$ are filled and separated from the empty at
$E_{1}(k)$ band by the energy $2m_{0}=E_{1}(\pi/3)-E_{2}(\pi/3)$.
Therefore, the second magnetization plateau appears at magnetization
equal to $2/3$ of the saturation value and the magnetic field at
 the center of plateau is given by
\begin{eqnarray}
H_{c2}^{0} &=& J^{0}_{\small AF}+E_{2}(\pi/3)+ m_{0}.
 \label{H-c2-0}
\end{eqnarray}
Since at finite band-filling the effect interaction between spinless
fermions cannot be ignored the width of plateau differs from its
bare value $2m_{0}$. In the forthcoming section we use the continuum
limit bosonization treatment of the effective spin-chain model
(\ref{Hamiltonian_SpFrm}) to determine parameters characterizing the
appearance and scale of the magnetization plateaus.

\subsection{Magnetization plateaus:  $H_{c_{1}}^{\pm}$ and \textcolor[rgb]{0.00,0.00,0.00}{$H_{c_{2}}^{\pm}$}}

To determine parameters characterizing the appearance and scale of
the magnetization plateaus, we use the continuum-limit bosonization
treatment of the model (\ref{effevtive-hamiltoni}). Following the
usual procedure in the low energy limit, we bosonize the spin
degrees of freedom at fixed magnetization $m$ and the interaction
term becomes~\cite{Cabra99}
\bea \tau_{n}^{z}  &=&  m + \sqrt{\frac{K}{\pi}} \partial_x \phi(x)
+ \frac{A_1}{\pi} \sin\left( \sqrt{4\pi K}\phi(x) +(2m+1)\pi n \right) \, ,\label{bosforTau_z}\\
\tau_{n}^{+} &=&e^{-i\sqrt{\pi/K}\theta(x)}\left[1+
\frac{B_1}{\pi}\sin\left(\sqrt{4\pi K}\phi(x)+(2m+1)\pi
n\right)\right]\, , \label{bosforTau_+} \eea
where $A_1$ and $B_1$ are  non-universal real constants of the
order of unity \cite{Hikihara01} and $m$ is the magnetization (per
site) of the chain. Here $\phi(x)$ and $\theta(x)$ are dual bosonic
fields, $\partial_t \phi = v_{s}
\partial_x \theta $, and satisfy the following commutation
relations
\begin{eqnarray}
\label{regcom}
&& [\phi(x),\theta(y)]  = i\Theta (y-x)\,,  \nonumber\\
&& [\phi(x),\theta(x)]  =i/2\, ,
\end{eqnarray}
and $K(\Delta,m)$ is the spin-stiffness parameter for chain with
anisotropy $\Delta$ and magnetization $m$. At zero magnetization
\begin{equation}
K(\Delta,0)=\frac{\pi}{2\left(1-\arccos\Delta \right)} \, .\label{K}
\end{equation}
Therefore, at $m=0$, $K=0.75$ for $J>0$ and $K=1.5$
for $J<0$. At the transition line into the ferromagnetic phase,
where the magnetization reaches its saturation value $m_{sat}=0.5$,
the spin stiffness parameter takes the universal value $K(\Delta,
0.5)=1$ \cite{Cabra98}. Respectively for finite magnetization, at
$0<m<m_{sat}$ and  $J>0$, the function $K(\Delta,m)$ monotonically
increases with enhancing $m$ and reaches its maximum value at
saturation magnetization, $K(\Delta, m_{sat})=1$, while for $J<0$,
monotonically decreases with increasing $m$ and reaches its minimum
value at saturation magnetization $K(\Delta, 0.5)=1$.

Using (\ref{bosforTau_z})\textcolor[rgb]{0.00,0.00,0.00}  {and}
(\ref{bosforTau_+}), in the case of "A" type of exchange
modulation, we get the following bosonized Hamiltonian
\begin{eqnarray}
 &H_{Bos}^{A}=\int dx \Big\{ \frac{v_{s}}{2}[(\partial_{x}\phi)^{2} +
(\partial_x\theta)^{2} ]- \textcolor[rgb]{0.00,0.00,0.00}{h_{eff}^{0}} \sqrt{\frac{K}{\pi}}\partial_{x}\phi +\nonumber\\
&\hspace{-15mm}+ \frac{h_{1}^{A}}{2\pi a_{0}}
\Big[\sin\left(2\pi(m+\frac{1}{6})n + \sqrt{4\pi K}\phi\right)
+
\sin\left(2\pi(m+\frac{5}{6})n+\sqrt{4\pi K}\phi\right)\Big] \Big\},
\label{Bosonized_Hamiltonian_A}
\end{eqnarray}
while in the case of "B" type of modulation - following
\begin{eqnarray}
&H_{Bos}^{B} = \int dx \Big\{
\frac{v_{s}}{2}[(\partial_{x}\phi)^{2} + (\partial_x\theta)^{2} ]-
\textcolor[rgb]{0.00,0.00,0.00}{h_{eff}^{0}}
\sqrt{\frac{K}{\pi}}\partial_{x}\phi+&\nonumber\\
&\hspace{-15mm}+ \frac{h_{1}^{B}}{2\pi a_{0}}\Big[
\cos\left(2\pi(m+\frac{1}{6})n + \sqrt{4\pi K}\phi\right)
-
\cos\left(2\pi(m+\frac{5}{6})n+\sqrt{4\pi K}\phi\right)\Big] \Big\}.
& \label{Bosonized_Hamiltonian_B}
\end{eqnarray}

From the bosonized Hamiltonian,
one easily gets the commensurate values of magnetization $m=\pm
1/6$ where a magnetization plateau appears. Away from this
commensurate values of magnetization, arguments of cosine terms
are strongly oscillating and in the continuum limit, their
contribution is irrelevant. Therefore in this case, the model is
described by the effective gaussian model, indicating on gapless
character of excitations and continuously increasing magnetization
of the chain with increasing magnetic field.

At commensurate  values of magnetization $m=\pm 1/6$, the cosine
term in (\ref{Bosonized_Hamiltonian_A}) and
(\ref{Bosonized_Hamiltonian_B}) is not oscillating and therefore
the modulated magnetic field term comes into play. Up to an
irrelevant shift on constant of the bosonic field, the generalized
Hamiltonian which  describes system at commensurate magnetization
is given by
\be \hspace{-20mm}H_{Bos}^{i}=\int dx \Big\{
\frac{v_{s}}{2}[(\partial_{x}\phi)^{2} + (\partial_x\theta)^{2} ]  +
\frac{m_{0}^{i}}{2\pi a_{0}} \cos(\sqrt{4\pi K}\phi)
 - h_{0} \sqrt{\frac{K}{\pi}}\partial_{x}\phi\Big\} \,\,  i=A,B.
\label{Bos_Hamiltonian-Gen} \ee
where
\be\label{Bare-Mass} m_{0}^{A}=\delta J_{AF}^{0} \qquad {\mbox and}
 \qquad m_{0}^{B}=\textcolor[rgb]{0.00,0.00,0.00}{2}m_{0}^{A}/\sqrt{3}\, .\ee

The Hamiltonian (\ref{Bos_Hamiltonian-Gen}) is the standard
Hamiltonian for the commensurate-incommensurate transition which
has been intensively studied in the past using
bosonization~\cite{Japaridze78} and the Bethe
ansatz~\cite{Japaridze84}. Below, we use these results to describe
the magnetization plateau and the transitions from a gapped
(plateau) to gapless paramagnetic phases.

Let us first consider
$\textcolor[rgb]{0.00,0.00,0.00}{h_{eff}^{0}=0}$. In this case, the
continuum theory of the initial hexamer spin-chain model in the
magnetic field $H=J_{\small AF}^{0} + J/4$ is given by the quantum
sine-Gordon (SG) model with a massive term $\sim
h_{1}^{i}\cos(\sqrt{4\pi K}\phi)$. From the exact solution of the
SG model~\cite{Dashen75}, it is known that the excitation spectrum
is gapless for $K \geq 2$ and has a gap in the interval $0 < K <
2$.  The exact relation between the soliton mass $M$ and the bare
mass $m_{0}$ is given by~\cite{Zamolodchikov95}
\be M = J{\cal C}(K) \left(m_{0}/J\right)^{1/(2-K)} \, ,
\label{SG-mass_Zamolodchikov} \ee
where
\bea \label{C-K} {\cal C}(K)\!&\!=\!&\! \frac{2}{\sqrt{\pi}}
\frac{\Gamma(\frac{K}{8-2K})}{\Gamma(\frac{2}{4-K})}
\left[\frac{\Gamma(1-K/4)}{2\Gamma(K/4)}\right]^{2/(4-K)}\!\!. \eea
Here $\Gamma$ is the Gamma function and the spin stiffness parameter
$K$ depends on the sign of $J$ and magnetization $m$.

In the gapped phase, the ground state properties of the system are
determined by the dominant potential energy term $\sim
\cos(\sqrt{4\pi K}\phi)$ and therefore in the gapped phases, the
field $\phi$ is pinned in one of the vacua
\be \la 0| \sqrt{4\pi K}\phi|0 \ra = (2n+1)\pi \, , \label{minima}
\ee
to ensure the minimum of the energy.

At $h_{eff}^{0} \neq 0$ (i.e. $H \neq J_{AF}^{0} + J/4$), the
presence of the gradient term in the Hamiltonian
(\ref{Bos_Hamiltonian-Gen}) makes it necessary to consider the
ground state of the sine-Gordon model in sectors with nonzero
topological charge. The effective chemical potential $\sim
h_{eff}^{0}\sqrt{\frac{K}{\pi}}\partial_{x}\phi$ tends to change
the number of particles in the ground state i.e. to create finite
and uniform density of solitons. It is clear that the gradient
term in (\ref{Bos_Hamiltonian-Gen}) can be eliminated by a gauge
transformation $\phi \rightarrow \phi +
h_{eff}^{0}\sqrt{\frac{K}{\pi}}\,x$, however this immediately
implies that the vacuum distribution of the field $\phi$ will be
shifted with respect of the minima (\ref{minima}). This
competition between contributions of the smooth and modulated
components of the magnetic field is resolved as a continuous phase
transition from a gapped state at $|h_{eff}^{0}| < M$ to a gapless
(paramagnetic) phase at $|h_{eff}^{0}| > M$, where $M$ is the
soliton mass~\cite{Japaridze78}.  This condition gives two
critical values of the magnetic field for each plateau
$H_{c,j}^{\pm} = H_{c,j}^{0}  \pm M_{j}$.

In the case of chain with AF exchange on weak bonds ($J>0$) this
gives
\bea H_{c,1}^{\pm} &=& J^{0}_{\small AF}  \pm J{\cal C}(K_{1})
\left(c_{i}\textcolor[rgb]{0.00,0.00,0.00}{\gamma}\right)^{1/(2-K_{1})}\, ,\\
H_{c,2}^{\pm} &=& J^{0}_{\small AF} + J \pm J{\cal C}(K_{2})
\left(c_{i}\textcolor[rgb]{0.00,0.00,0.00}{\gamma}\right)^{1/(2-K_{2})}\,
, i=A,B \label{Hc1-2-pm} \eea
where
\bea K_{1} &=&K(1/2,-1/6)\, , \quad {\mbox and } \quad K_{2}
=K(1/2,+1/6)\, .  \label{K-1-2-AF} \eea
and $c_{A}=1$,
$c_{B}=\textcolor[rgb]{0.00,0.00,0.00}{2}/\sqrt{3}$.

Respectively, in the case of chain with ferromagnetic exchange on
weak bonds ($J<0$) we obtain
\bea H_{c,1}^{\pm} &=& J^{0}_{\small AF} + J \pm |J|{\cal
C}(K^{\prime}_{1})
\left(c_{i}\textcolor[rgb]{0.00,0.00,0.00}{\gamma}\right)^{1/(2-K^{\prime}_{1})}\, ,\\
H_{c,2}^{\pm} &=& J^{0}_{\small AF}  \pm |J|{\cal
C}(K^{\prime}_{2})
\left(c_{i}\textcolor[rgb]{0.00,0.00,0.00}{\gamma}\right)^{1/(2-K^{\prime}_{2})}\,
, i=A,B \label{Hc1-2-pm} \eea
where
\bea K^{\prime}_{1} &=&K(-1/2,-1/6)\, , \quad {\mbox and } \quad
K^{\prime}_{2} =K(-1/2,+1/6)\, .  \label{K-1-2-AF} \eea

Correspondingly the width of each magnetization plateaus is given by
\be \hspace{-17mm}H_{c,j}^{+}- H_{c,j}^{-} = 2M_{j} = \left\{
\begin{array}{l@{\quad}}2J{\cal C}(K_{j}) \left(m_{0}^{j}/J
\right)^{1/(2-K_{j})} \hspace{9mm} {\mbox at} \quad J>0 \\
2|J|{\cal C}(K^{\prime}_{j}) \left(m_{0}^{j}/|J|
\right)^{1/(2-K^{\prime}_{j})} \quad{\mbox at} \quad
J<0\end{array}\right. \, j=1,2. \label{Platou} \ee

To estimate the numerical value of the spin stiffness parameter
$K$ at magnetization $m$ and anisotropy $\Delta$ we use the
following {\it \textcolor[rgb]{0.00,0.00,0.00}{ansatz}}
\be
K(\Delta,m)=K(\Delta,0)+\frac{|m|}{m_{sat}}\left(1-K(\Delta,0)\right)\,
.\label{Anzatz}
 \ee
Here, we assume that with increasing magnetization, the spin
stiffness parameter {\em monotonically reaches its extremum $K=1$
at} $m=m_{sat}$. This ansatz gives that $K_{1}=K_{2} \simeq 0.87$
for $J>0$  and $K_{1}=K_{2}\simeq 1.335$ for $J<0$. It is
straightforward to get, that in the antiferromagnetic case the
width of the magnetization plateau scales as $\delta^{1.13}$ while
\textcolor[rgb]{0.00,0.00,0.00}{in the case of chain with
ferromagnetic weak bonds }as $\delta^{1.50}$.

In order to investigate the detailed behavior of the ground state
magnetic phase diagram and to test the validity of the picture
obtained from continuum-limit bosonization treatment, below in
this paper we present results of numerical calculations using the
Lanczos method of exact diagonalization for finite chains.
\begin{figure}
\centerline{\psfig{file=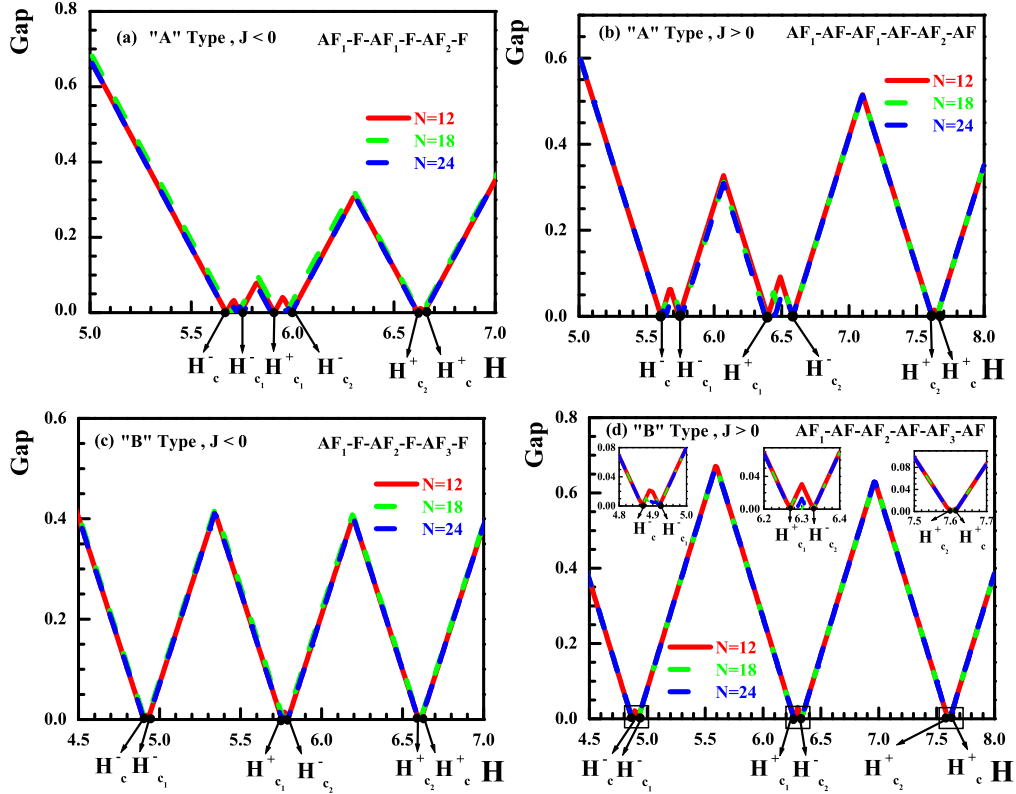,width=6.00in}}
\caption{(Color online.) Difference
between the energy of the first lowest level and the ground state
(Gap) as a function of the magnetic field, $H$ for chains with
exchanges (a)~$J_{AF}^{0}=\frac{19}{3}$, $\delta=\frac{2}{19}$ and
$J=-1$, (b)~$J_{AF}^{0}=\frac{19}{3}$, $\delta=\frac{2}{19}$ and
$J=1$, (c)~$J_{AF}^{0}=6$, $\delta=\frac{1}{6}$ and
$J=-1$ , (d)~$J_{AF}^{0}=6$, $\delta=\frac{1}{6}$ and
$J=1$ and lengths $N=12,18$ and $24$.} \label{Gap}
\end{figure}

\section{Numerical simulation} \label{sec4}

A very famous and accurate method in field of numerical simulation
is known as the Lanczos method. In order to explore the nature of
the spectrum and the phase transition, we use the Lanczos method
to diagonalize numerically the model (\ref{Hamiltonian}) with
periodic boundary conditions.

In this section, to find the effect of a magnetic field on the
ground state phase diagram of the model, we calculate the spin
gap, the magnetization, the string correlation function and the
bond-dimer order parameter for finite chains with lengths $N=12, 18, 24$ and periodic boundary conditions.

Since in a critical field, the energy gap should be closed, the
best way to find the critical fields is the investigation of the
energy gap, which is recognized as the difference between the
energies of the first exited state and the ground state in finite
chains. In Fig.~(\ref{Gap}), we have presented results of these
calculations on the energy gap for the exchanges parameter
corresponding to the "A" and "B" types as
\begin{eqnarray}
&(a)&~~ J_{AF}^{0}=\frac{19}{3},~~ \delta=\frac{2}{19},~~  J=-1, \nonumber \\
&(b)&~~ J_{AF}^{0}=\frac{19}{3},~~ \delta=\frac{2}{19},~~  J=+1,\nonumber \\
&(c)&~~ J_{AF}^{0}=6,~~~~ \delta=\frac{1}{6},~~ ~~   J=-1,\nonumber\\
&(d)&~~ J_{AF}^{0}=6,~~~~ \delta=\frac{1}{6},~~~~ J=+1.
\end{eqnarray}
and chain lengths $N=12,18, 24$. At $H=0$, it is clearly seen that
the spectrum of the model is gapped. As soon as the magnetic field
applies, the energy gap decreases linearly with $H$ and vanishes
at $H_{c}^{-}$. This is the first level crossing between the
ground state energy and the first excited state energy. By further
increasing the magnetic field, three gapless regions
\begin{eqnarray}
H_{c}^{-}&<& H <H_{c_{1}}^{-}, \nonumber \\
H_{c_{1}}^{+}&<& H <H_{c_{2}}^{-},\nonumber\\
H_{c_{2}}^{+}&<& H <H_{c}^{+},
\end{eqnarray}
and three gapped regions
\begin{eqnarray}
H_{c_{1}}^{-}&<& H <H_{c_{1}}^{+}, \nonumber \\
H_{c_{2}}^{-}&<& H <H_{c_{2}}^{+},\nonumber\\
H&>&H_{c}^{+}.
\end{eqnarray}
 are seen in Fig.~\ref{Gap}.  Oscillations of the energy gap in gapless regions are the result of level crossings in
finite size systems. One should note that the energy gap in the last gapped region, $H>H_{c}^{+}$, growths linearly with the magnetic field $H$ which is an indication of the saturated ferromagnetic phase. Therefore in respect to the non modulated case $\delta=0$ \cite{mahdavifar08}, two new gapped regions are created by adding $\delta$.
\begin{figure}
\centerline{\psfig{file=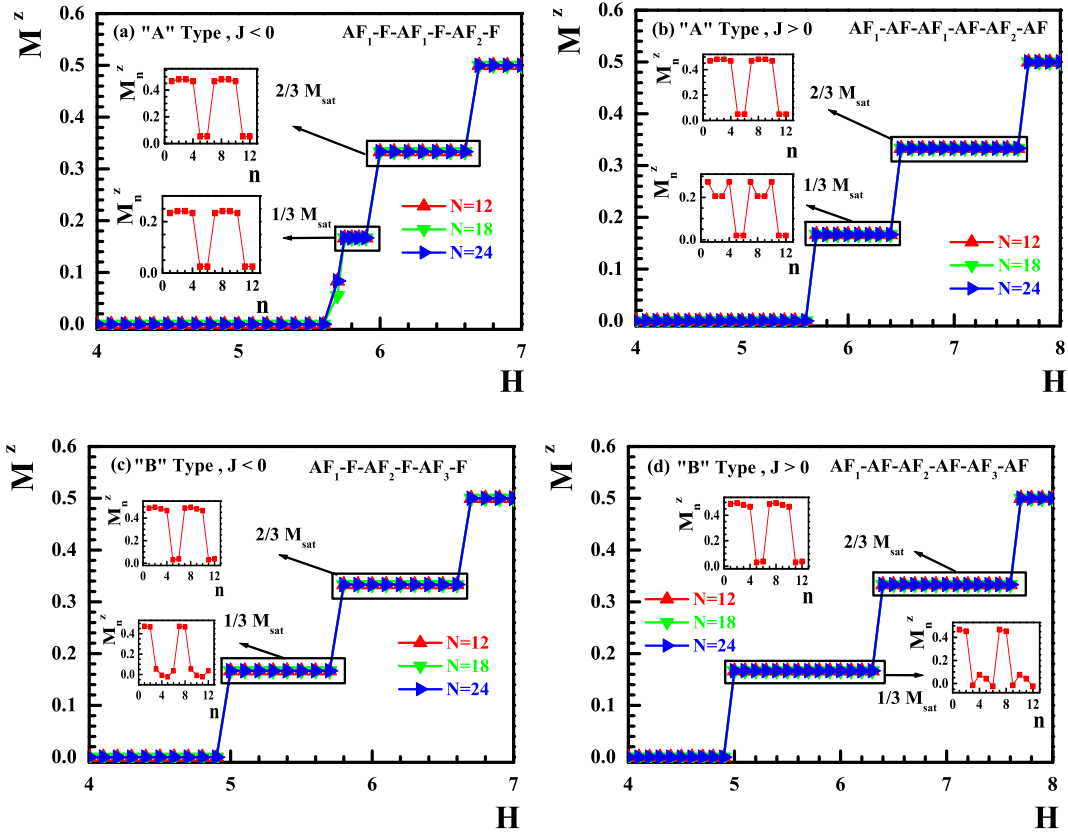,width=6.30in}}
\caption{(Color online.) Magnetization as a function of the magnetic field, $H$ for chains with exchanges (a)~$J_{AF}^{0}=\frac{19}{3}$,
 $\delta=\frac{2}{19}$ and $J=-1$, (b)~$J_{AF}^{0}=\frac{19}{3}$, $\delta=\frac{2}{19}$ and $J=1$, (c)~$J_{AF}^{0}=6$, $\delta=\frac{1}{6}$ and
$J=-1$ , (d)~$J_{AF}^{0}=6$, $\delta=\frac{1}{6}$ and $J=1$  and
lengths $N=12,18$ and $24$. In the insets , the magnetization on
site as a function of the site number "n" is plotted for a value
of the magnetic field in the region of plateau of $1/3M_{sat}$ and
$2/3M_{sat}$ for length $N=24$.}~~~~~~~~~~~~~~~~~~~~~~~~
\label{Magnetization}
\end{figure}
To find the critical fields, we have used the phenomenological
renormalization group (PRG) method \cite{mahdavifar08}. As an
example, critical fields for the exchanges
$J_{AF}^{0}=\frac{19}{3}$, $\delta=\frac{2}{19}$, $J=-1$ are given
as follows:
\begin{eqnarray}
H_{c}^{-}&=&5.67\pm 0.01,~~~~~~~~~~~H_{c}^{+}=6.65\pm 0.01,\nonumber  \\
H_{c_{1}}^{-}&=&\textcolor[rgb]{0.00,0.00,0.00}{5.75}\pm 0.01,~~~~~~~~~~~H_{c_{1}}^{+}=5.91\pm0.01, \nonumber \\
H_{c_{2}}^{-}&=&6.00\pm 0.01,~~~~~~~~~~~H_{c_{2}}^{+}=6.62\pm0.01\label{critical-fields}
\end{eqnarray}

\begin{figure}
\centerline{\psfig{file=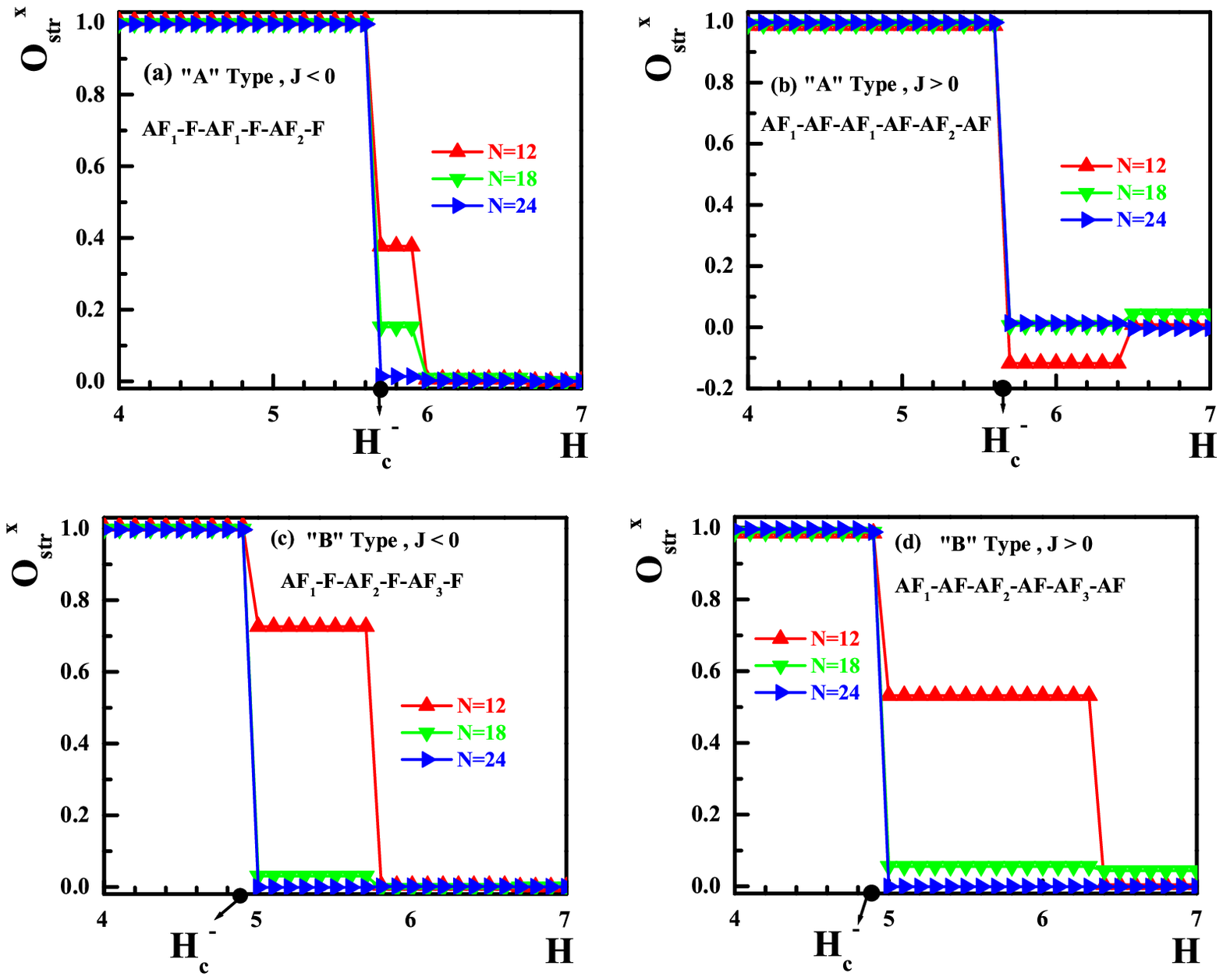,width=6.00in}}
 \caption{(Color online.) String correlation function as a function of the magnetic field, $H$ for chains with exchanges
(a)~$J_{AF}^{0}=\frac{19}{3}$, $\delta=\frac{2}{19}$ and
$J=-1$, (b)~$J_{AF}^{0}=\frac{19}{3}$, $\delta=\frac{2}{19}$ and
$J=1$, (c)~$J_{AF}^{0}=6$, $\delta=\frac{1}{6}$ and
$J=-1$ , (d)~$J_{AF}^{0}=6$, $\delta=\frac{1}{6}$ and
$J=1$ and
lengths $N=12$, $18$ and $24$.} \label{String order}
\end{figure}

The first insight into the magnetic order of the ground state of
the system determined by studying the magnetization process.  The
magnetization along the field axis is defined as
\begin{eqnarray}
M^{z}=\frac{1}{N}\sum_{n=1}^{N}\langle
Gs|S_{n}^{z}|Gs \rangle, \label{magnetization}
\end{eqnarray}
where the notation $\langle Gs|...|Gs\rangle$ represents the
ground state expectation value. In Fig.~\ref{Magnetization}, we
have plotted the magnetization along the applied magnetic field,
$M^{z}$, vs $H$ for chain lengths $N=12, 18, 24$ and different
exchange parameters corresponding to the "A" and "B" types. As it
is clearly seen, in absence of the magnetic field $H=0$, the
magnetization is zero. By increasing the magnetic field, the
magnetization remains zero up to the first critical field
$H=H_{c}^{-}$. This part of the magnetization is known as the
zero-plateau. This behavior is in agreement with expectation based
on general statement that in the gapped phases, the magnetization
along the applied field appears only at a finite critical value of
the magnetic field equal to the spin gap. Besides the standard
zero and saturation plateaus at $H<H_{c}^{-}$ and $H>H_{c}^{+}$
respectively, two additional plateaus are seen at
$M^{z}=\frac{1}{3}M_{sat}$ and $M^{z}=\frac{2}{3}M_{sat}$, where
both of them satisfy the Oshikawa-Yamanaka-Affleck $(OYA)$
condition. To check that the middle plateaus are not finite size
effect, we performed the size scaling \cite{zhitomirsky00} of
their width and found that the size of the plateaus interpolates
to finite value when $N\longrightarrow\infty$. As it has clearly
seen in Fig.~\ref{Gap} and Fig.~\ref{Magnetization}, width of
plateaus and the mentioned gapped regions  grow by increasing the
modulation $\delta$. In the insets of Fig.~\ref{Magnetization},
the magnetization on site, $M^{z}_{n}=\langle
Gs|S^{z}_{n}|Gs\rangle$, as a function of the site number $n$ is
plotted for some values of the magnetic field corresponding to the
plateaus at $\frac{1}{3}M_{sat}$ and $\frac{2}{3}M_{sat}$ and
length $N=24$. As we observe, the system shows a well pronounced
modulation of the on site magnetization, where magnetization on
weak modulated-bonds is larger than on strong modulated-bonds.
This distribution remains almost unchanged within the plateau at
$\frac{1}{3}M_{sat}$ for $H_{c_{1}}^{-}< H < H_{c_{1}}^{-}$ and
the plateau at $\frac{2}{3}M_{sat}$ for fields $H_{c_{2}}^{-} < H
< H_{c_{2}}^{+}$.

By analyzing the results on the energy gap, we found that the
spectrum is gapped in absence of the magnetic field which is one
of the properties of the Haldane phase. since, we have considered the very strong
antiferromagnetic (AF) exchange on odd bonds, the Haldane phase cannot be found and it is expected to be the  gapped dimer phase. This phase can be
recognized from studying the string correlation function. The string
correlation function in a chain of length $N$, defined only for
odd $l$, is~\cite{hida99}
\begin{eqnarray}
O_{Str}(l, N)=-\langle exp\{i\pi \sum_{2n+1}^{2n+l+1} S_{k}^{x}\}\rangle.
\label{string}
\end{eqnarray}

In particular, we calculated the string correlation function for
different finite chain lengths $N=12, 18, 24$.    Since the model has the $U(1)$ symmetry in presence of a magnetic field, we consider the transverse component of the string correlation
function. Fig.~\ref{String order} presents the Lanczos results on
the string correlation function for the chain with lengths $N=12, 18,
24$ and different exchange parameters corresponding to the "A" and
"B" types. As can be seen from this figure, at
$H<H_{c}^{-}$,
the string correlation function $O_{Str}(l, N)$ is saturated and
the hexamer chain system is in the dimer phase. As soon as the
magnetic field increases from the first critical filed, as
expected the string correlation function decreases very rapidly and reaches zero in the thermodynamic limit,
which shows that only the dimer phase remains stable in the
presence of a magnetic field less than $H_{c}^{-}$. The nonzero values of the string correlation
function in the region $H>H_{c}^{-}$, are finite size effects and  is completely clear that in the thermodynamic limit $N \longrightarrow \infty$ will be zero.

\begin{figure}
\centerline{\psfig{file=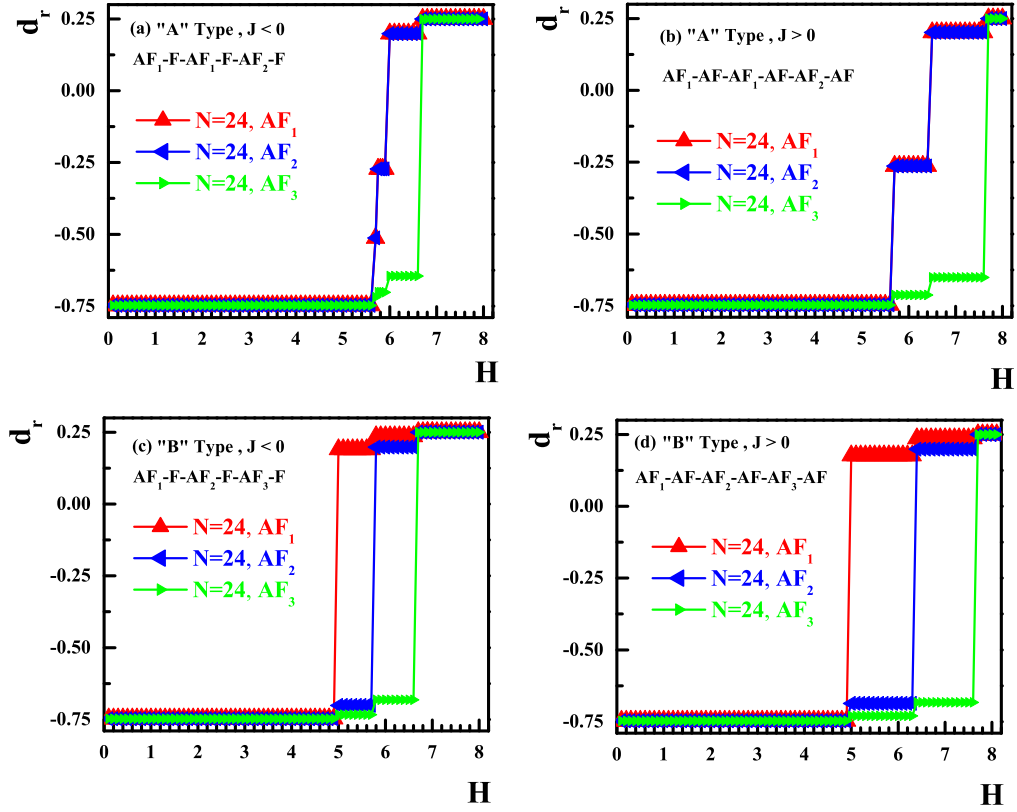,width=6.00in}}
 \caption{(Color online.) The bond-dimer order parameter as a function of the magnetic field, $H$ for chains with exchanges (a)~$J_{AF}^{0}=\frac{19}{3}$, $\delta=\frac{2}{19}$ and
$J=-1$, (b)~$J_{AF}^{0}=\frac{19}{3}$, $\delta=\frac{2}{19}$ and
$J=1$, (c)~$J_{AF}^{0}=6$, $\delta=\frac{1}{6}$ and
$J=-1$ , (d)~$J_{AF}^{0}=6$, $\delta=\frac{1}{6}$ and
$J=1$ and
length $N=24$.}\label{dimer}
\end{figure}
An additional insight into the nature of different phases can be
obtained by studying the correlation functions. Since there are three kind of the space-modulated exchanges in our hexameric chain  model, we define the following  bond-dimer order parameters;

\begin{eqnarray}
d_{r}^{AF_{1}}&=&\frac{6}{N}\sum_{n=0}^{\textcolor[rgb]{0.00,0.00,0.00}{(N/6)-1}}<Gs|\textbf{S}_{6n+1}.\textbf{S}_{6n+2}|Gs>,\nonumber
\\
d_{r}^{AF_{2}}&=&\frac{6}{N}\sum_{n=0}^{\textcolor[rgb]{0.00,0.00,0.00}{(N/6)-1}}<Gs|\textbf{S}_{6n+3}.\textbf{S}_{6n+4}|Gs>,\nonumber
\\
d_{r}^{AF_{3}}&=&\frac{6}{N}\sum_{n=0}^{\textcolor[rgb]{0.00,0.00,0.00}{(N/6)-1}}<Gs|\textbf{S}_{6n+5}.\textbf{S}_{6n+6}|Gs>,
\end{eqnarray}
where summations are taken over the space-modulated
antiferromagnetic bonds. In
Fig.~\ref{dimer}\textcolor[rgb]{0.00,0.00,0.00}{,} we have plotted
$d_{r}^{AF_{1}}$, $d_{r}^{AF_{2}}$ and $d_{r}^{AF_{3}}$ versus
magnetic field $H$ for chain of length $N=24$ with different
exchange parameters corresponding to the "A" and "B" types. As it
is seen from this figure, at $H<H_{c}^{-}$ spins on all
antiferromagnetic space-modulated bonds are in a singlet state
$d_{r}^{AF_{1}}=d_{r}^{AF_{2}}=d_{r}^{AF_{3}} \simeq -0.75$, while
at $H>H_{c}^{+}$, \textcolor[rgb]{0.00,0.00,0.00}{the bond-dimer
order parameter} $d_{r}$, is equal to the saturation value
$d_{r}^{AF_{1}}=d_{r}^{AF_{2}}=d_{r}^{AF_{3}}\sim 0.25$ and the
ferromagnetic long-range order along the magnetic field axis is
present. However, in the considered case of strong
antiferromagnetic exchanges ($J_{AF}^{0}\gg |J|$) and high
critical fields, quantum fluctuations are substantially suppressed
and calculated averages of  spin correlations are very close to
their nominal values.

For intermediate values of the magnetic field, at
$H_{c}^{-}<H<H_{c}^{+}$ the data presented in Fig.~\ref{dimer}
gives us a possibility to trace the mechanism of singlet-pair
melting with increasing the magnetic field. As it follows from
Fig.~\ref{dimer}, at values of the magnetic field slightly above
$H_{c}^{-}$ spin singlet pairs start to melt in all modulated
bonds simultaneously and almost with the same intensity. By
further increasing  of $H$ and for fields $H>H_{c}^{-}$, melting
of weak modulated bonds gets more intensive, however at
$H=H_{c_{1}}^{-}$ the process of melting stops and the bond-dimer
order parameter remains constant up to the critical field
$H=H_{c_{1}}^{+}$. As it is seen in Fig.~\ref{dimer}(a) and (b)
for the "A" type of the modulation, in the $1/3$-plateau state,
weak and strong modulated bonds manifest on-site singlet features
with dimerization values $\simeq -0.25$ and $\simeq -0.70$,
respectively. In contrast, for the "B" type of the modulation
(Fig.~\ref{dimer}(c) and (d)), weak modulated bonds are almost
polarized with dimerization value $\simeq 0.18$, while
 intermediate and strong bonds
manifest strong on-site singlet features with value $\simeq
-0.70$. By more increasing the magnetic field and for fields
$H>H_{c_{1}}^{+}$, the melting of the weak bonds happens
intensively and at $H=H_{c_{2}}^{-}$ the process of melting stops
and remains stable up to the critical field $H_{c_{2}}^{+}$. It is
clearly seen, in the $2/3$-plateau state, weak bonds for "A" type
and weak and intermediate modulated bonds for "B" type are
polarized  ($\simeq 0.22$), but strong bonds still manifest strong
on-site singlet features ($\simeq -0.65$). Finally, for
$H_{c_{2}}^{+}$ strong bonds start to melt more intensively while
the polarization of weak bonds increases slowly and at
$H=H_{c}^{+}$ all weak and strong bonds achieve an identical,
almost fully polarized state.

\section{Conclusion}

In this paper, we have  studied the effect of additional
modulation of strong antiferromagnetic bonds on the ground state
magnetic phase diagram of the 1D spin-1/2 chain with alternating
exchange.  In particular, we focus our studies on the
chain with hexamer modulation, where the spin exchange on even bonds
is the same, while the strong antiferromagnetic exchange on odd
bonds is modulated with period three.

In the limit where the odd couplings are dominant, we mapped the
model to an effective XXZ Heisenberg chain with anisotropy
$\Delta$ in an effective uniform and spatially trimer modulated
magnetic field. The anisotropy parameter is $\Delta=\frac{1}{2}$
in the case of antiferromagnetic exchange on even bonds and
$\Delta=-\frac{1}{2}$ in the case of ferromagnetic exchange on
even bonds. Using the continuum-limit bosonization treatment of
the effective spin-chain model, we have shown that the additional
modulation of the strong bonds with period three and amplitude
$\simeq \delta$ leads to generation of two gaps in the excitation
spectrum of the system at magnetization equal to the 1/3 and 2/3
of its saturation value. As a result of this new energy scale
formation, the magnetization curve of the system $M(H)$ exhibits
two plateaus at $M=\frac{1}{3}M_{sat}$ and $M=\frac{2}{3}M_{sat}$.
The width of the plateaus, is proportional to the excitation gap
and scales as $\delta^{\nu}$, where critical exponent
$\textcolor[rgb]{0.00,0.00,0.00}{\nu =1.13\pm0.01}$ in the case
of a AF exchange on even bonds and $\nu =1.50\pm0.01$ in the case of
ferromagnetic exchange on even bonds.

For complete description of the model, we supplement our analysis by a very accurate numerical
simulation. Using the Lanczos method of numerical diagonalizations
for chains up to $N=24$,  we have studied the effects of an
external magnetic field on the ground state properties of the
system. In the first part of the numerical experiment, we have
focused on the energy gap of the system. Our results showed that,
in respect to the non-modulated chain, two new gapped regions
create by adding the modulation. The widths of the mentioned
gapped regions grow by increasing the parameter of modulation
$\delta$. In the second part of the numerical experiment, we have
studied the magnetization process. We provided a clear picture of
the magnetization which showed that two magnetization plateaus
appear at values $\frac{1}{3}M_{sat}$ and $\frac{2}{3}M_{sat}$ in
the new gapped regions. To find additional insight into the nature
of different phases, we also calculated the on-site magnetization,
the string correlationfunction and the bond-dimer order parameters.
The on-site magnetization showed a microscopic picture of the
direction of spins on different sites, when system is in the new
gapped regions.
 On the
other hand, by studying the string correlation function, we found that
in the absent of the magnetic field, the suggested alternating
chain is in the dimer phase and this phase remains stable in the
presence of an external magnetic field up to the first critical
field. Finally, we studied the effect of the magnetic filed on the
ground state phase diagram of the model, by means of the
perturbation approach. Using perturbation theory, we provided the
analytical results for the critical fields that these results were
in well agreement with the obtained numerical experiment results.

\section*{Acknowledgments}

We wish to thank  N. Avalishvili for a helpful communication. GIJ
acknowledges support from the SCOPES Grant IZ73Z0-128058 and the
Georgian NSF Grant No. ST09/4-447.

\section*{Appendix}

\section*{Perturbation results}

In this section, we study the effect of the magnetic filed on the ground state phase diagram of the model, using the perturbation approach.  We try to find the analytical results for the critical fields. The behavior of the model (\ref{Hamiltonian}) in the limit of strong couplings on the odd bonds $J_{AF}(n)\gg J$ is interested.
For this aim, it is convenient to
rewrite the Hamiltonian (\ref{Hamiltonian}) in the form
\begin{eqnarray}
{\cal H} &=& {\cal H}_{0}+{\cal V}\nonumber \\
\nonumber \\
{\cal H}_{0} &=& \sum_{n=1}^{N/2}J_{AF}(n) \textbf{S}_{2n-1}.\textbf{S}_{2n}-h \sum_{n=1}^{N} S_n^{z}
\nonumber \\
{\cal V}&=& J\sum_{n=1}^{N/2}\textbf{S}_{2n}.\textbf{S}_{2n+1}.  \label{Per-Hamiltonian}
\end{eqnarray}

The unperturbed part, ${\cal H}_{0}$, is the Hamiltonian of $N/2$
non-interacting
 pairs of spins. The eigenstate of the unperturbed Hamiltonian is written as a product of pair states.
 By solving eigenvalue equation of
an individual pair, one can easily find the eigenstates as
mentioned in Sec.~III. Let us start with the case of $\delta=0$.
Since the ground state energy of a distinct pair is two fold
degenerate at $H =J_{AF}(n)$ , the ground state energy of
unperturbed Hamiltonian ${\cal H}_{0}$ is $2^{N/2}$ times
degenerate ~\cite{mila98}. The perturbation ${\cal V}$ splits this
degeneracy. By applying the first order and second order
perturbation theory for finite chains with periodic boundary
conditions and generalize results to the thermodynamic limit, one
can find two critical fields. In the case modulated chain,
$\delta\neq 0$, there are different kinds of bonds: strong and
weak. In this case, by increasing the magnetic field, first weak
bonds melt and go to the triplet state in respect to the strong
bonds. Therefore, it is naturally to find four additional critical
fields in respect to the non-modulated case, $\delta=0$.

The determined critical fields for Hexameric chain of the "A" type with $J<0$ and by using the perturbation approach are
~~~~~~~~~~~~~~~~~~~~~~~~~~~~~~~~~~~~~

\begin{table}
\begin{center}
\label{table1} \caption{The critical fields which are obtained by
the perturbation theory and the corresponding values obtained by
the numerical experiment in the Hexameric chain of the "A" type
and $J<0$, the "A" type and $J>0$, the "B" type and $J<0$ and the
"B" type and $J>0$. ~~~~~~~~~~~~~~~~~~~~~~~~~~ }
~~~~~~~~~~~~~~~~~~~~~~~~~

~~~~~~~~~~~~~~~~\begin{tabular}
{c} \hline \hline \hline \\
Critical fields~~~~~~~~~~Perturbation Results~~~~~~~Numerical Results\\ \\
\hline \hline \ "A" type and $J<0$\\
\hline \hline \\
$H_{c}^{-}$~~~~~~~~~~~~~~~~~~~~~~$5.62$~~~~~~~~~~~~~~~~~~~~~~~~$5.67\pm 0.01$\\

$H_{c_{1}}^{-}$~~~~~~~~~~~~~~~~~~~~~~$5.75$~~~~~~~~~~~~~~~~~~~~~~~~$5.75\pm 0.01$\\

$H_{c_{1}}^{+}$~~~~~~~~~~~~~~~~~~~~~~$5.87$~~~~~~~~~~~~~~~~~~~~~~~~$5.91\pm 0.01$\\

$H_{c_{2}}^{-}$~~~~~~~~~~~~~~~~~~~~~~$6.00$~~~~~~~~~~~~~~~~~~~~~~~~$6.00\pm 0.01$\\

$H_{c_{2}}^{+}$~~~~~~~~~~~~~~~~~~~~~~$6.62$~~~~~~~~~~~~~~~~~~~~~~~~$6.62\pm0.01$\\

$H_{c}^{+}$~~~~~~~~~~~~~~~~~~~~~~$6.62$~~~~~~~~~~~~~~~~~~~~~~~~$6.65\pm0.01$\\ \\
\hline \hline \ "A" type and $J>0$\\
\hline \hline \\
$H_{c}^{-}$~~~~~~~~~~~~~~~~~~~~~~~~~$5.62$~~~~~~~~~~~~~~~~~~~~~~~~$5.61\pm 0.01$\\

$H_{c_{1}}^{-}$~~~~~~~~~~~~~~~~~~~~~~~~~$5.75$~~~~~~~~~~~~~~~~~~~~~~~~$5.75\pm 0.01$\\

$H_{c_{1}}^{+}$~~~~~~~~~~~~~~~~~~~~~~~~~$6.37$~~~~~~~~~~~~~~~~~~~~~~~~$6.40\pm 0.01$\\

$H_{c_{2}}^{-}$~~~~~~~~~~~~~~~~~~~~~~~~~$6.50$~~~~~~~~~~~~~~~~~~~~~~~~$6.59\pm 0.01$\\

$H_{c_{2}}^{+}$~~~~~~~~~~~~~~~~~~~~~~~~~$7.62$~~~~~~~~~~~~~~~~~~~~~~~~$7.61\pm0.01$\\

$H_{c}^{+}$~~~~~~~~~~~~~~~~~~~~~~~~~$7.62$~~~~~~~~~~~~~~~~~~~~~~~~$7.66\pm0.01$\\ \\
\hline \hline \ "B" type and $J<0$  \\
\hline \hline \\
$H_{c}^{-}$~~~~~~~~~~~~~~~~~~~~~~~~~$4.90$~~~~~~~~~~~~~~~~~~~~~~~~$4.90\pm 0.01$\\

$H_{c_{1}}^{-}$~~~~~~~~~~~~~~~~~~~~~~~~~$4.90$~~~~~~~~~~~~~~~~~~~~~~~~$4.96\pm 0.01$\\

$H_{c_{1}}^{+}$~~~~~~~~~~~~~~~~~~~~~~~~~$5.75$~~~~~~~~~~~~~~~~~~~~~~~~$5.75\pm 0.01$\\

$H_{c_{2}}^{-}$~~~~~~~~~~~~~~~~~~~~~~~~~$5.75$~~~~~~~~~~~~~~~~~~~~~~~~$5.78\pm 0.01$\\

$H_{c_{2}}^{+}$~~~~~~~~~~~~~~~~~~~~~~~~~$6.59$~~~~~~~~~~~~~~~~~~~~~~~~$6.60\pm0.01$\\

$H_{c}^{+}$~~~~~~~~~~~~~~~~~~~~~~~~~$6.59$~~~~~~~~~~~~~~~~~~~~~~~~$6.61\pm0.01$\\ \\
\hline \hline \ "B" type and $J>0$  \\
\hline \hline \\
$H_{c}^{-}$~~~~~~~~~~~~~~~~~~~~~~~~~$4.90$~~~~~~~~~~~~~~~~~~~~~~~~$4.87\pm 0.01$\\

$H_{c_{1}}^{-}$~~~~~~~~~~~~~~~~~~~~~~~~~$4.90$~~~~~~~~~~~~~~~~~~~~~~~~$4.92\pm 0.01$\\

$H_{c_{1}}^{+}$~~~~~~~~~~~~~~~~~~~~~~~~~$6.25$~~~~~~~~~~~~~~~~~~~~~~~~$6.27\pm 0.01$\\

$H_{c_{2}}^{-}$~~~~~~~~~~~~~~~~~~~~~~~~~$6.25$~~~~~~~~~~~~~~~~~~~~~~~~$6.33\pm 0.01$\\

$H_{c_{2}}^{+}$~~~~~~~~~~~~~~~~~~~~~~~~~$7.59$~~~~~~~~~~~~~~~~~~~~~~~~$7.60\pm0.01$\\

$H_{c}^{+}$~~~~~~~~~~~~~~~~~~~~~~~~~$7.59$~~~~~~~~~~~~~~~~~~~~~~~~$7.61\pm0.01$\\

\hline \hline \hline \\

\end{tabular}
\end{center}
\end{table}
\begin{eqnarray}
H_{c}^{-}&=&J_{AF}^{0}(1-\frac{\delta}{2})+\frac{J}{4}-\frac{J^ {2}}{12 \delta J_{AF}^{0}},\nonumber  \\
H_{c_{1}}^{-}&=&J_{AF}^{0}(1-\frac{\delta}{2})+\frac{J}{4},\nonumber  \\
H_{c_{1}}^{+}&=&J_{AF}^{0}(1-\frac{\delta}{2})-\frac{J^ {2}}{12 \delta J_{AF}^{0}},\nonumber  \\
H_{c_{2}}^{-}&=&J_{AF}^{0}(1-\frac{\delta}{2}),\nonumber  \\
H_{c_{2}}^{+}&=&J_{AF}^{0}(1+\frac{\delta}{2})+\frac{J}{4}+\frac{J^ {2}}{12 \delta J_{AF}^{0}},\nonumber  \\
H_{c}^{+}&=&J_{AF}^{0}(1+\frac{\delta}{2})+\frac{J}{4}+\frac{J^ {2}}{12 \delta J_{AF}^{0}}.\label{a.perturbation}
\end{eqnarray}
Generalization of this result for "A" type and $J>0$, "B" type and $J<0$ and "B" type and $J>0$ is straightforward.
In Table 1, the numerical and perturbation analytical results of critical fields were compared for  chain of the "A" type and $J<0$, the "A" type and $J>0$, the "B" type and $J<0$ and the "B" type and $J>0$. The accuracy of analytical results are to two significant digits. We emphasize that the perturbation results are in well agreement with the obtained numerical experiment results.

~~~~~~~~~~~~~~~~~~~~~~~~~~~~~~~~~~~~~~~~~~~~~~~~~~~~~~~~~~~~~
~~~~~~~~~~~~~~~~~~~~~~~~~~~~~~~~~~~~~~~~~~~~~~~~~~~~~~~~~~~~~
~~~~~~~~~~~~~~~~~~~~~~~~~~~~~~~~~~~~~~~~~~~~~~~~~~~~~~~~~~~~~




\end{document}